\documentstyle[aps,epsfig,float,twocolumn,prl]{revtex}
\newcommand{\be}{\begin{equation}}
\newcommand{\ee}{\end{equation}}
\newcommand{\bea}{\begin{eqnarray}}
\newcommand{\eea}{\end{eqnarray}}
\begin{document}
\twocolumn[\hsize\textwidth\columnwidth\hsize\csname @twocolumnfalse\endcsname
\draft
\title{Vibrational edge modes in intrinsically inhomogeneous doped transition metal oxides}
\author{I. Martin, R. J. McQueeney, and A. R. Bishop}
\address{
Los Alamos National Laboratory, Los Alamos, NM 87545}
\author{Z. G. Yu}
\address{SRI International, Menlo Park, CA 94025}
\date{\today}
\maketitle
\begin{abstract}
By applying an unrestriced Hartree-Fock approximation and a Random Phase
approximation to multiband Peierls-Hubbard Hamiltonians, we determine the
phonon mode structure in models of transition metal oxides in the presence of
intrinsic nanoscale inhomogeneities induced by hole doping.  We identify low
frequency $local$ vibrational modes pinned to the interface between regions of
distinct electronic structure and separated in frequency from the band of
extended phonons. A major characteristic of these ``edge'' modes is that their
energy is essentially insensitive to the doping level, while their intensity
increases with the density of interfaces (and thus doping level). We argue that
the presence of such modes is a typical feature of systems with phase
separation, including cuprates, nickelates, manganites, and bismuthates. We
specifically address the experimental signatures of these modes in lattice
inelastic neutron scattering.
\vskip2mm
\end{abstract}

] Recent advances in experimental techniques provide strong indications that a
broad class of electronic materials, commonly referred to as {\em strongly
correlated}, exhibit intrinsically inhomogeneous electronic
phases\cite{1,2,3,4,5,6,7}. Strong interactions in these materials drive phase
separation between regions of distinct electronic structure\cite{dagotto}.
However, a competing interaction or electronic kinetic energy can lead to
frustration of the global phase separation and to formation of {\em nanoscale
inhomogeneities}, which can take the form of, e.g., stripes\cite{zaanen} or
clumps\cite{koulakov}.  It has been realized early on that unlike the
conventional spin and charge density wave phases, the ordering of intrinsic
inhomogeneities is sensitive to thermal and quantum fluctuations\cite{fradkin},
as well as static disorder. This makes an unambiguous identification of such
phases, and their distinction from the more common density wave instabilities
problematic, and therefore requires development of new diagnostics, which are
more sensitive to the unique structure of such phases.  In  this Letter, we
propose one such diagnostic tool which relies on detection of localized {\em
edge modes} pinned to the interface between the regions of different electronic
structure.  While the edge (interface) modes are predicted to appear in every
degree of freedom coupled to the inhomogeneity (charge, spin,
lattice)\cite{yuPRB}, here we focus of the {\em lattice vibrational edge modes}
which lend themselves readily to study with inelastic neutron scattering.

We consider two particular classes of transition metal oxides that have
attracted significant attention recently -- cuprates and nickelates. The former
is the material exhibiting high-$T_c$ superconductivity, and the latter is a
related compound with Cu replaced by Ni. The Ni-based compounds are typically
insulating, with unambiguously demonstrated stripe order\cite{1}. In cuprates,
on the other hand, the presence of stripes and their possible influence on
superconductivity are subjects of intensive debate. Since it appears that
stripes can significantly affect superconductivity\cite{kivel,martin,bishop},
it is important to establish whether they are indeed present in the material.
Our purpose here is to elucidate typical manifestations of stripe order through
a comparative theoretical study of cuprates and nickelates.  In particular, we
compare these theoretical results to observations of anomalous phonon modes in
the cuprates\cite{robCu1,robCu2,ybco} and nickelates\cite{Ni,robNi2}. The
results also apply to edge mode signatures in other doped transition metal
oxides (and related complex electronic materials), as found in recent
experimental studies\cite{Ni,Bi,eCu,CMR}; equivalent modes have been predicted
and identified in conjugated polymers and quasi-1D (one-dimensional) charge
transfer solids\cite{Raman,RRaman}.

We first analyze the signatures of the edge modes in neutron scattering for
doped nickelates.  To model the NiO$_2$ planes, we use a 2D four band extended
Peierls-Hubbard Hamiltonian, which includes both electron-electron and
electron-phonon interactions\cite{yone,yi}:
\bea\label{eq:H0}
H_0 = \sum_{<ij>, m, n,
\sigma}{t_{im,jn}(u_{ij})(c^\dagger_{im\sigma}c_{jn\sigma}+ H.c.)} \\\nonumber+
\sum_{i,m,\sigma}{\epsilon_m(u_{ij})c^\dagger_{im\sigma}c_{im\sigma}}+\sum_{<ij>}{\frac12
K_{ij}u^2_{ij}} + H_c^{Ni}.
\eea
Here, $c^\dagger_{im\sigma}$ creates a hole with spin
$\sigma$ and site $i$ in orbital $m$ (there are two nickel orbitals
(Ni d$_{x^2-y^2}$ and d$_{3z^2-1}$) and two oxygen orbitals (O p$_x$ and
p$_y$)included).  The bare Ni-O hopping $t_{im,jn}$ has two values:
$t_{pd}$ between d$_{x^2-y^2}$ and p and $\pm t_{pd}/\sqrt{3}$ between
d$_{3z^2-1}$ and p.  The O site electronic energy is $\epsilon_p$ and the Ni
site energies are
$\epsilon_d$ and $\epsilon_d+E_z$, with $E_z$ the crystal-field splitting on
the Ni site.  The correlation part of the Hamiltonian is modelled by
\bea\label{eq:Hc}
H_c^{Ni} = \sum_{im}{(U+2J)n_{im\uparrow}n_{im\downarrow}} - \sum_{i, m\ne
n}{2J{\bf S}_{im}\cdot {\bf S}_{in}} \\\nonumber +
\sum_{i,m\ne n,\sigma, \sigma^\prime}{(U-J/2)n_{im\sigma}n_{in\sigma^\prime}} + \sum_{i,m,n}{J
c^\dagger_{im\uparrow}c^\dagger_{im\downarrow}c_{in\downarrow}c_{in\uparrow}}.
\eea
Here, $U$ is the on-site Ni Coulomb repulsion, $J$ is the Hund coupling that
favors a high spin state on Ni sites, and ${\bf S}_{im} =
\frac12\sum_{\alpha,\beta}{c^\dagger}_{im\tau}{\bf
\sigma}_{\tau\tau^\prime}c_{im\tau^\prime}$, with $\bf{\sigma}$ the vector of Pauli
matrices.  The electron-lattice interaction causes modification of the Ni-O
hopping strength through the oxygen displacement $u_{ij}$: $t_{im,jn}(u_{ij}) =
t_{im,jn}(1\pm
\alpha u_{ij})$, where $+(-)$ applies if the Ni-O bond shrinks (stretches) for a
positive $u_{ij}$; it also affects the Ni on-site energies
$\epsilon_m(u_{ij}) = \epsilon_d + \beta\sum_j{(\pm u_{ij})}$, where the sum runs over the
four neighboring O ions. The other oxygen modes couple to electron charge more
weakly, and are neglected here for simplicity.  In our calculations we use the
following set of parameters\cite{zaan,robNi}:
$\epsilon_p-\epsilon_d = 9, \ U =4,\ J = 1, \ E_z = 1$, and $K = 32/{\AA}$ (all in units of
$t_{pd}$; we take $t_{pd} = 1$ eV).  The electron-lattice couplings $\alpha = 3/\AA$ and $\beta = 1/\AA$ were varied
to best fit the neutron scattering data.  To approximately solve the model, we
use unrestricted Hartree-Fock (HF) combined with an inhomogeneous Random Phase
Approximation (RPA) for linear lattice fluctuations\cite{yone,robNi} in a
supercell of size
$N_x\times N_y$ with periodic boundary conditions.  In this model,
doped holes tend to segregate into the stripes due to competition between
magneto-elastic interaction that favors global electronic phase separation and
electronic kinetic energy that favors uniform carrier density. It is worth
noting that other competing interactions can produce stripes, clumps and other
inhomogeneities\cite{kiv,branko}, but the local coupled charge-spin-lattice
dynamics governing edge modes on these templates can still be modelled with the
above Hamiltonian and RPA analysis.

The output of the calculation is the inhomogeneous HF groundstate and
the phonon eigenfrequencies and eigenvectors.  From the phonon
eigenmodes, we calculate the corresponding neutron scattering cross
section, defined as \bea S({\bf k},\omega) =\int{dt\, e^{-i\omega
    t}\sum_{l l^\prime}{\langle e^{-i{\bf k R}_l(0)}e^{i{\bf k
        R}_{l^\prime}(t)}\rangle}}, \eea where ${\bf R}_{l}(t) = {\bf
  R}_{l}^0 + {\bf d}_l + {\bf u}_l(t)$ is the position of the $l$-th
oxygen atom expressed in terms of the location of the unit cell origin
${\bf R}_{l}^0$, position within the unit cell ${\bf d}_l$, and
time-dependent vibrational component ${\bf u}_l(t)$.  For phonon modes
with ${\bf u}_l(t)$ oriented along the corresponding metal-oxygen
bonds, on the O$_x$ sublattice ${\bf d}_l = \frac a2 \hat{x}$ and
${\bf u}_l\equiv x_l \hat{x}$, and on the O$_y$ sublattice ${\bf d}_l
= \frac a2 \hat{y}$ and ${\bf u}_l\equiv y_l \hat{y}$.  The scalar
displacements can now be expressed in terms of the normal modes $z_n$
as $x_l(t) = \sum_n{\alpha_{x_l,n}z_n(t)}$ and $y_l(t) =
\sum_n{\alpha_{y_l,n}z_n(t)}$.  Making first order expansion in the
oxygen displacements, we obtain \bea S({\bf k},\omega) &=&
\sum_n{\Big[k_x^2|\alpha_{{\bf k},n}^x|^2+k_y^2|\alpha_{{\bf
      k},n}^y|^2}\\\nonumber && + k_x
k_y(e^{i(k_x-k_y)a/2}\alpha_{{\bf k},n}^x \alpha_{-{\bf k},n}^y +
c.c.)  \Big]\\\nonumber &&\times\frac{\hbar}{2m\omega_n}
[(1+n_B)\delta(\omega-\omega_n) + n_B\delta(\omega+\omega_n)].  \eea
Here, $\alpha_{{\bf k},n}^x = \sum_l{e^{-i{\bf k
      R}_l^0}\alpha_{x_l,n}}$, and $n_B = (e^{\omega_n/T}-1)^{-1}$ is
the thermal population of the phonon mode $n$. This is a
generalization of the usual neutron scattering intensity
expression\cite{lovesay} for the case of phonons with a larger real
space unit cell.  In this paper we plot $S({\bf k},\omega)/|{\bf
  k}|^2$ for {\bf k}-directions sampling longitudinal modes,
consistent with the common experimental convention.

\vspace{-.2cm}
\begin{figure}[htbp]
\begin{center}
\includegraphics[width = 1.4 in]{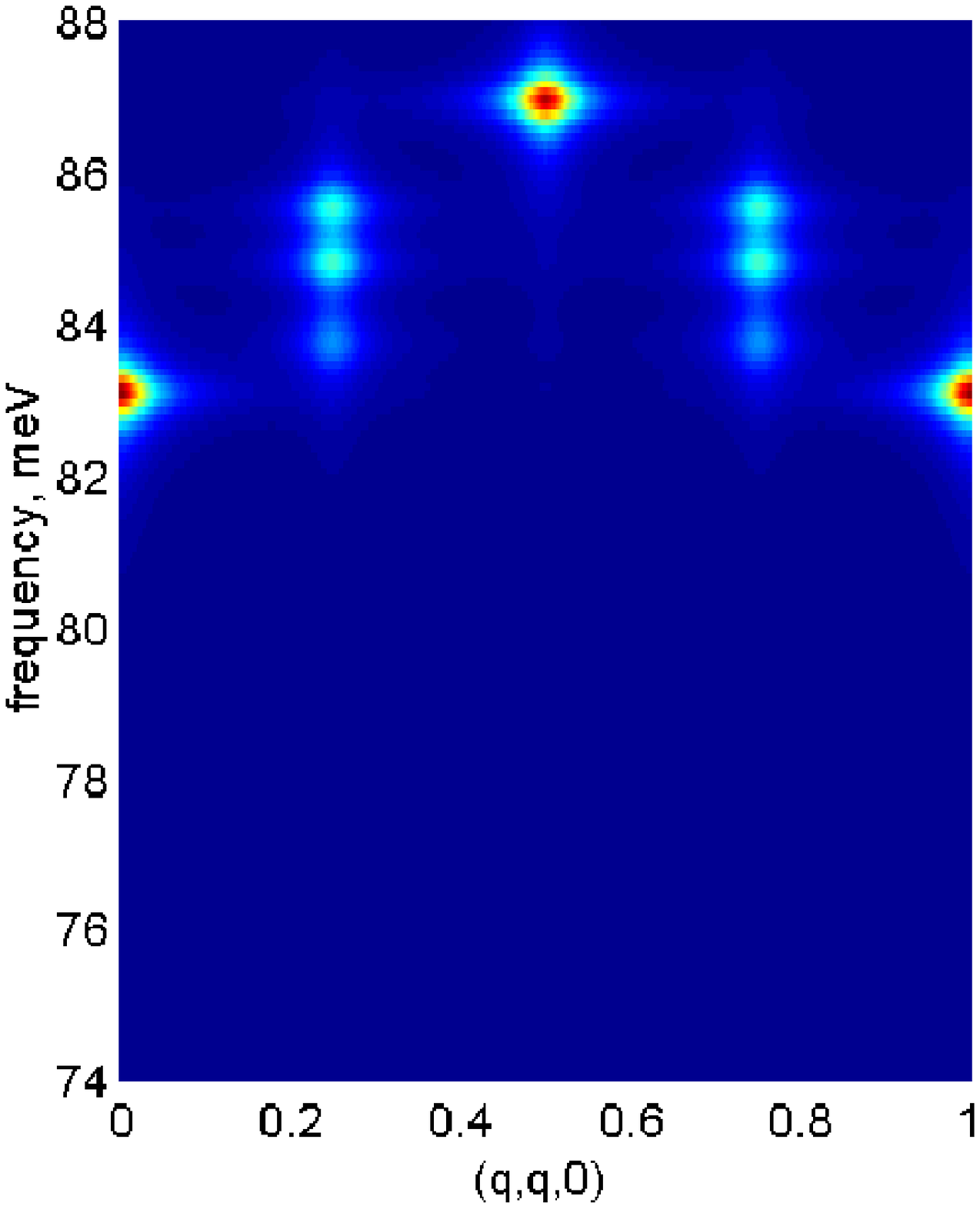}
\includegraphics[width = 1.4 in]{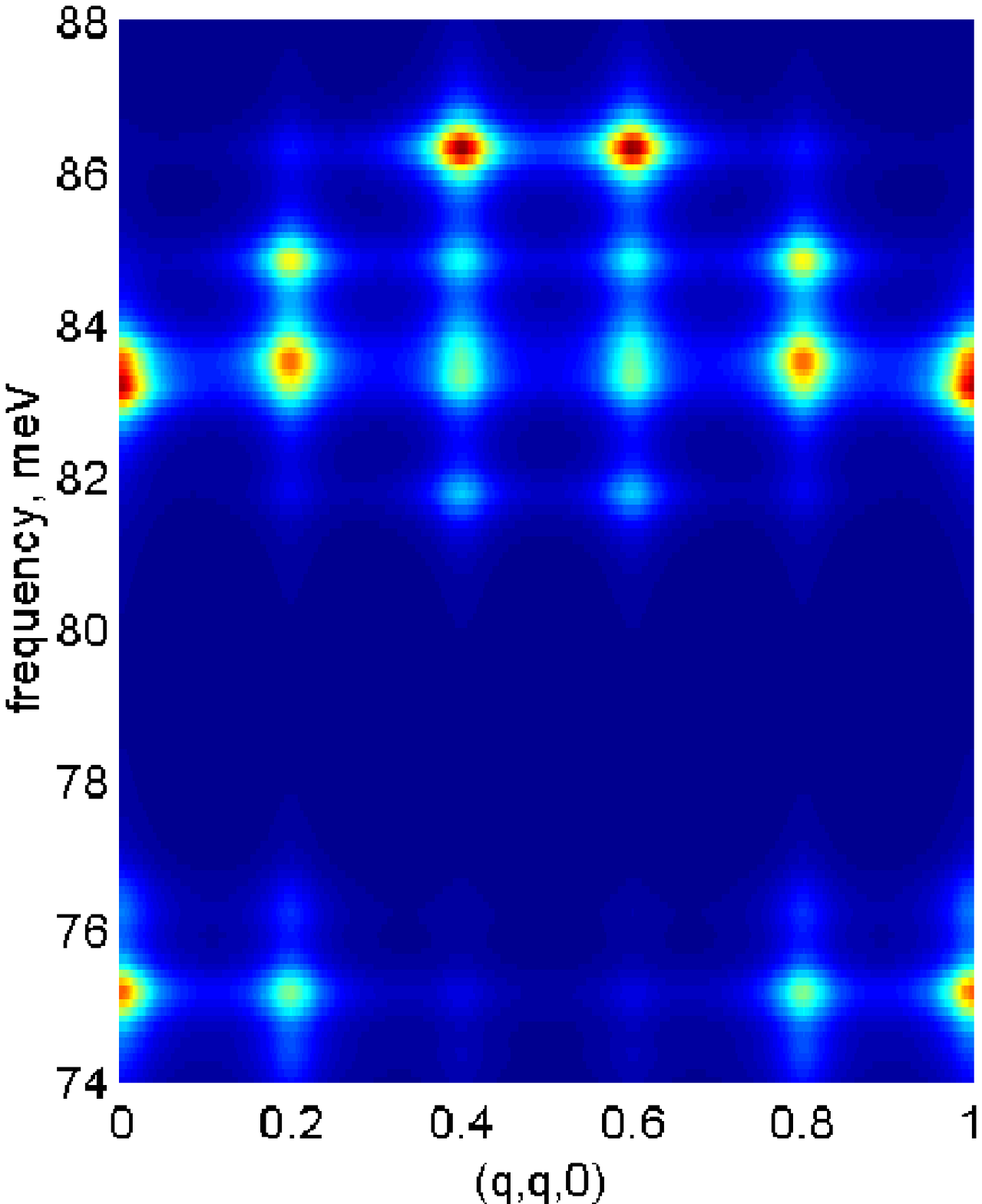}\\
\includegraphics[width = 1.4 in]{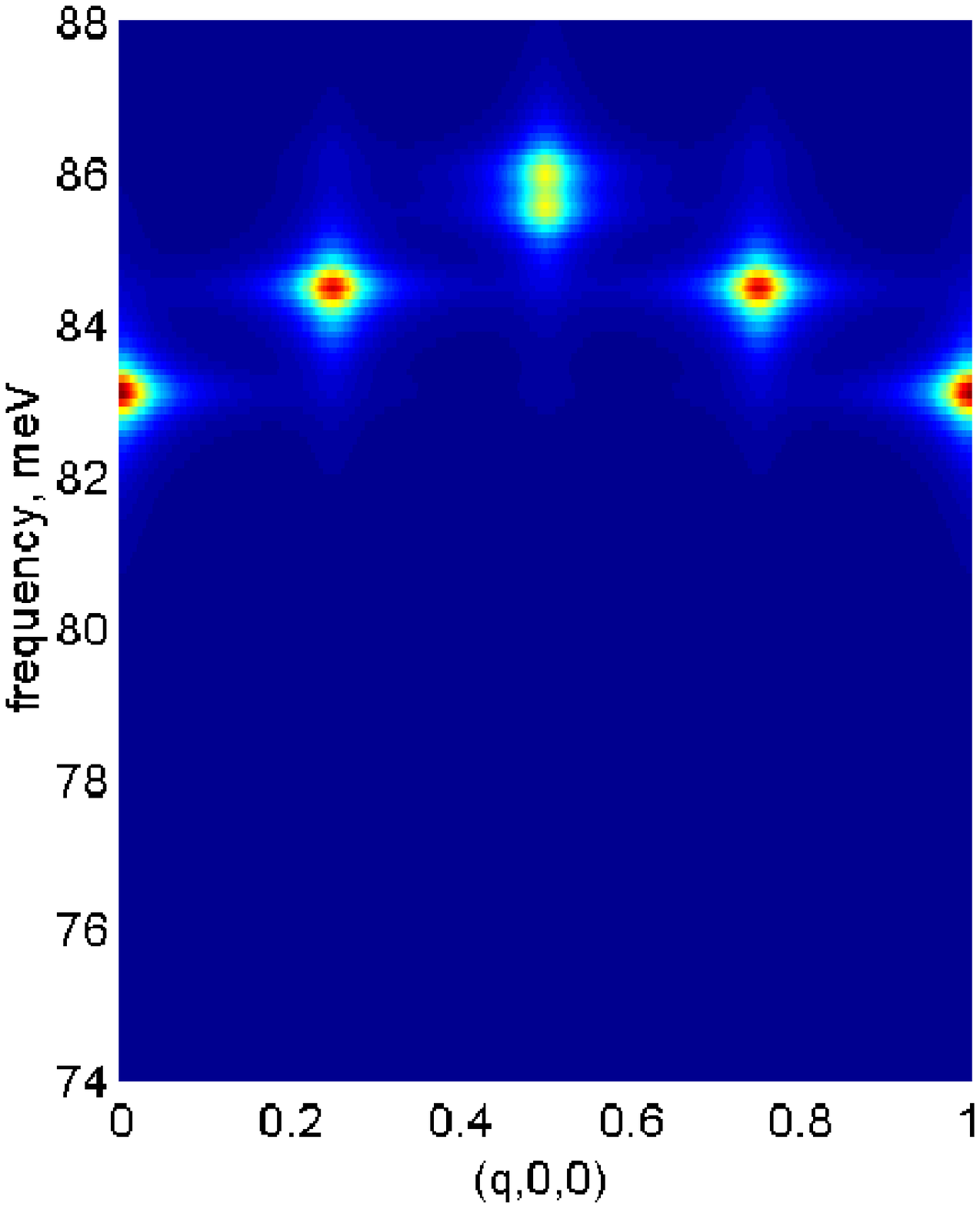}
\includegraphics[width = 1.4 in]{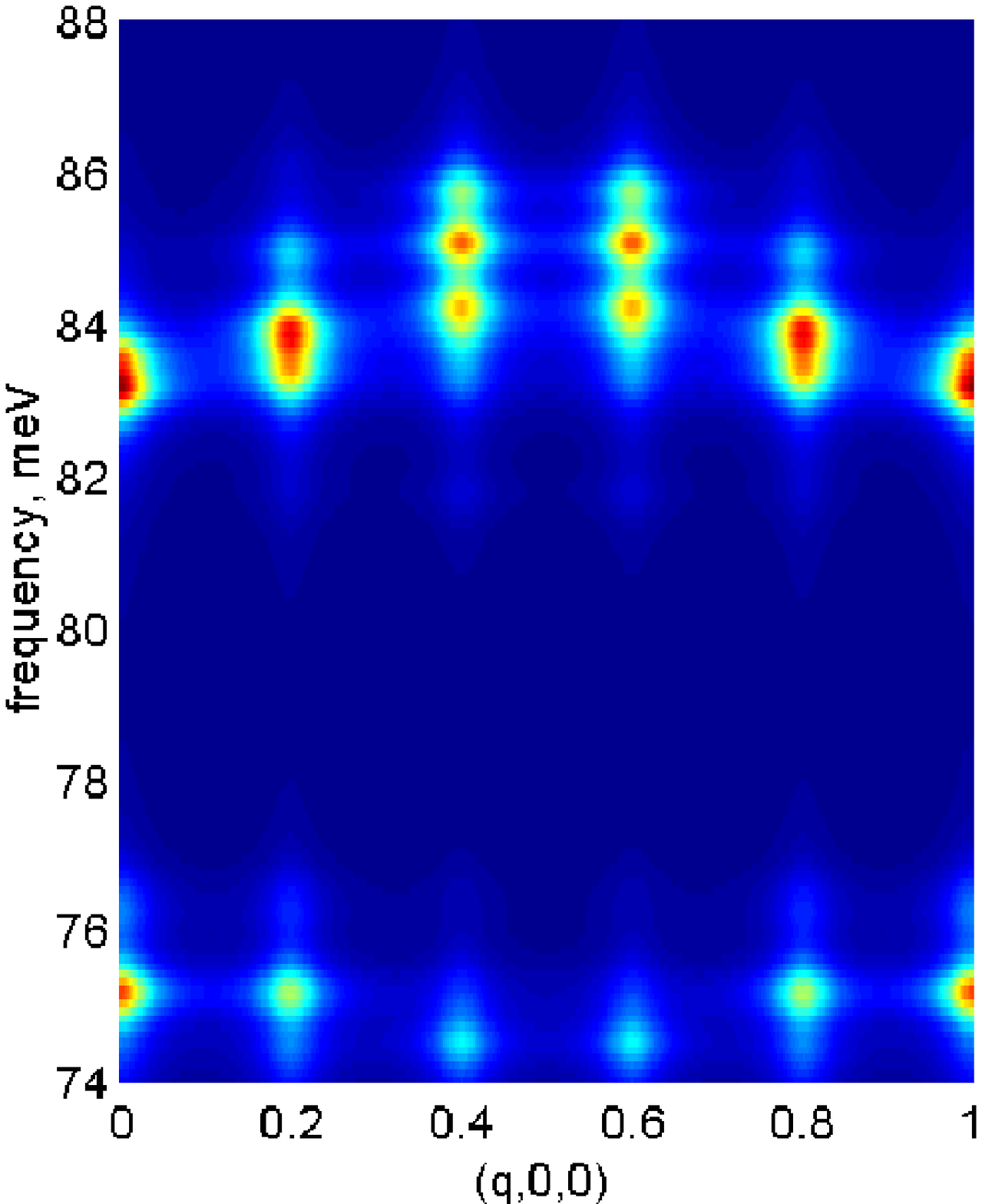}\\
\hspace{.8 cm}\includegraphics[width = .96 in]{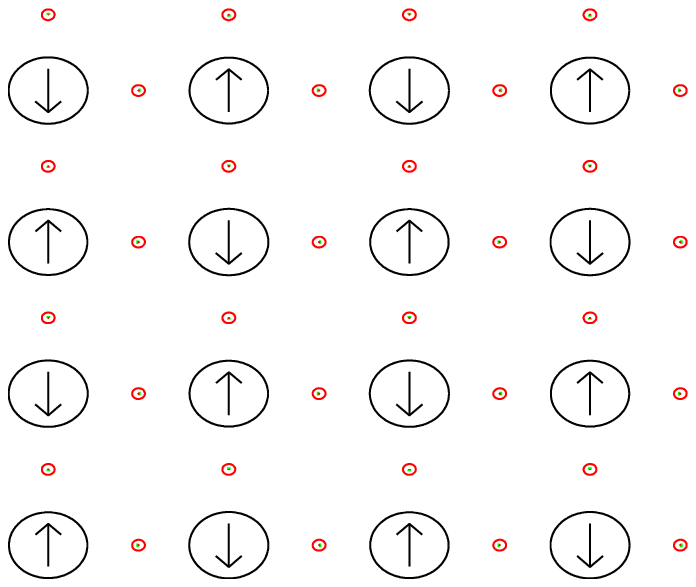}\hspace{.9 cm}
\includegraphics[width = 1.2 in]{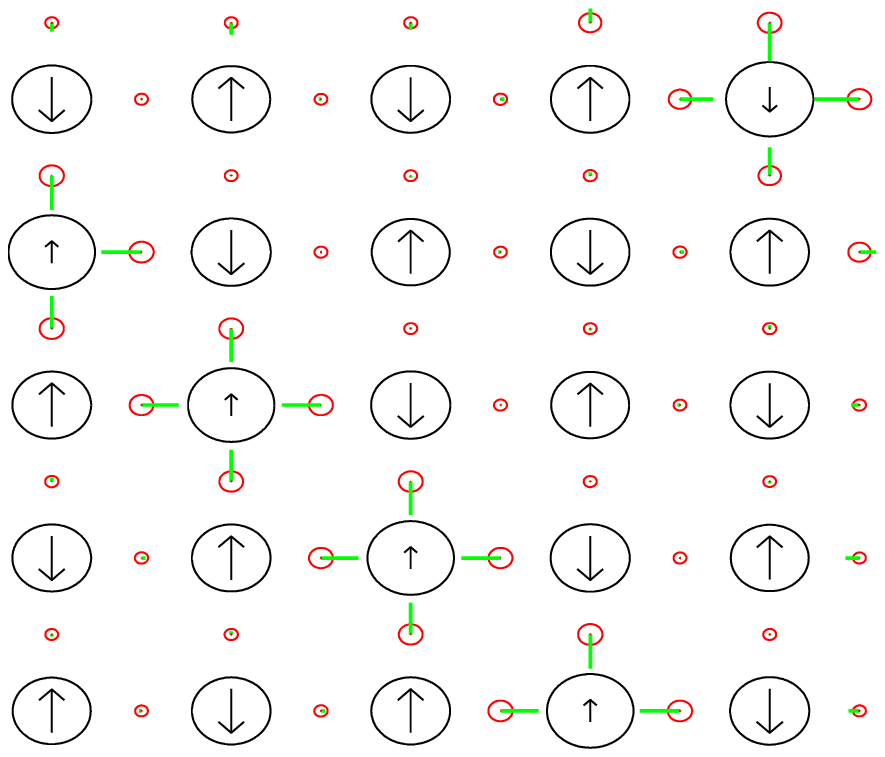}
\vspace{0.4cm} \caption{Neutron scattering spectra and the ground
  state structures for the undoped (left panels), and hole-doped ($x =
  0.2$) NiO$_2$ planes.  The neutron scattering spectra are
  symmetrized with respect to directions along and across the stripe;
  in all plots $q$ runs from 0 to $2\pi$. The relative intensity color scheme
  is defined in Fig.~\ref{fig:Cu}. In the lower plots, black
  circles represent Ni sites and red circles represent O. The radius
  of a circle is proportional to the corresponding site hole density.
  Arrows centered on circles show the magnitude and direction of spin.
  The green lines originating from O sites indicate the magnitude and
  direction of the equilibrium O displacements due to the formation of
  a stripe.}
\label{fig:Ni}
\end{center}
\end{figure}

\vspace{-.4cm}
The ground state configurations and the corresponding neutron scattering
intensities for NiO$_2$ planes are shown in Fig.~\ref{fig:Ni}.  The ground
state configuration in the undoped case is a simple antiferromagnet (AF),
while, in the doped ($x = 0.2$) case, the holes spatially segregate into a Ni
centered stripe along the diagonal of the NiO$_2$ plane.  The neutron
scattering intensities are given in the original Brillouin zone, symmetrized
with respect to the orientation of the stripe.

The most striking new feature in the dynamic structure factor that appears upon
doping is the creation of the low frequency modes ($\omega \sim 75$ meV) that
split off from the main branch.  These low frequency modes have been observed
experimentally with inelastic neutron scattering \cite{Ni,robNi2}.  We find
that with increasing doping (e.g., $x = 0.5$) the spectral weight in the new
branch increases. However, its energy does not significantly change, nor is
there a dependence of the spectrum on the stripe spacing.  Intensity to create
the local modes is taken from the extended phonon mode branch, with
accompanying phase shifts of the extended phonons. There is no simple
dispersion that can be associated with the local modes and their wave vectors
are related to the characteristic $spatial$ $extent$ of the eigenmodes. This is
consistent with the doping insensitivity of the phonon ``anomaly'' observed in
neutron scattering experiments for nickelates reported in Ref.~\cite{robNi2}.
The presence of the phonon edge mode with doping-independent energy is
similarly consistent with the doping-insensitivity of the ``kink'' energy in
Angle Resolved Photoemission Spectra (ARPES) reported recently\cite{lanz} for
cuprates (discussed later).

\begin{figure}[htbp]
\begin{center}
\includegraphics[width = 1.35 in]{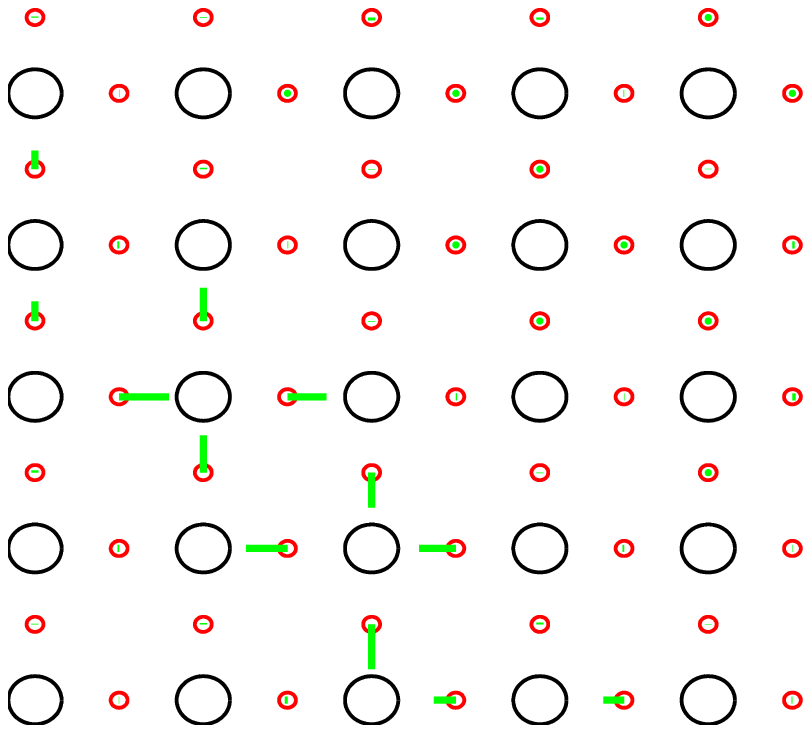}\hspace{.5cm}
\includegraphics[width = 1.4 in]{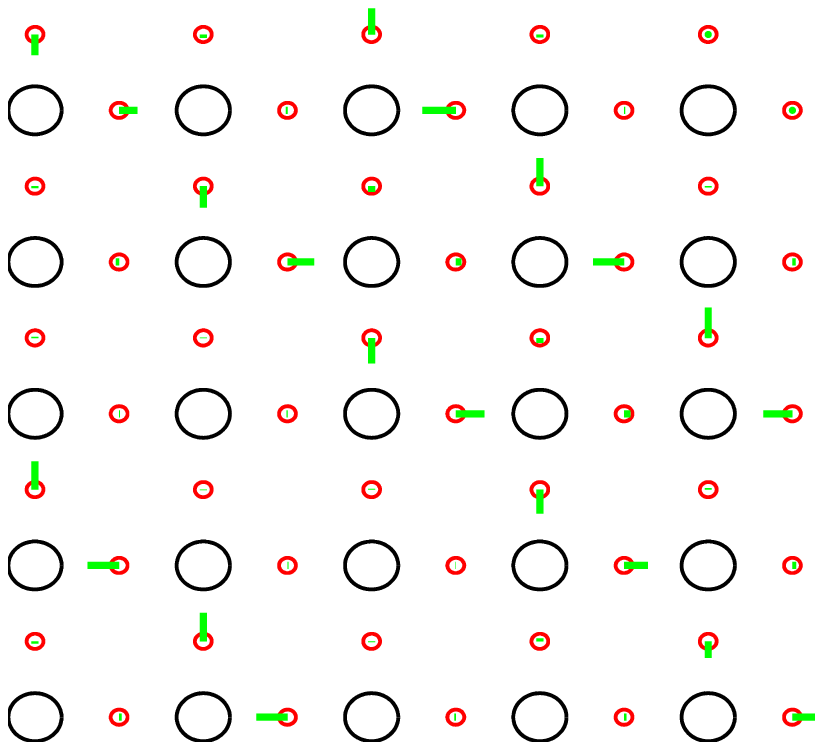} \vspace{0.4cm}
\caption{Representative vibrational eigenmodes for the case of a $x=0.2$
doped NiO$_2$ plane.  Black and red circles represent the equilibrium positions
of Ni and O atoms, respectively, and green lines show the magnitude and
direction of the O displacement in the mode. On the left, a low frequency mode
localized in the vicinity of the stripe ($E = 75.1$ meV); on the right, an
extended phonon with energy $E = 83.4$ meV. }
\label{fig:Ni_mode}
\end{center}
\end{figure}

In order to elucidate the structure of these new modes, in
Fig.~\ref{fig:Ni_mode} (left panel) we plot a representative mode from this
branch.  Clearly, this is a mode localized around the stripe edge.  For
comparison, in the right panel we plot a mode from the high energy branch.
Although modified (phase-shifted) due to the presence of a stripe, this mode
preserves its original spatially extended structure, characteristic of the
undoped case.

\vspace{-.2cm}

\begin{figure}[htbp]
\begin{center}
\includegraphics[width = 1.4 in]{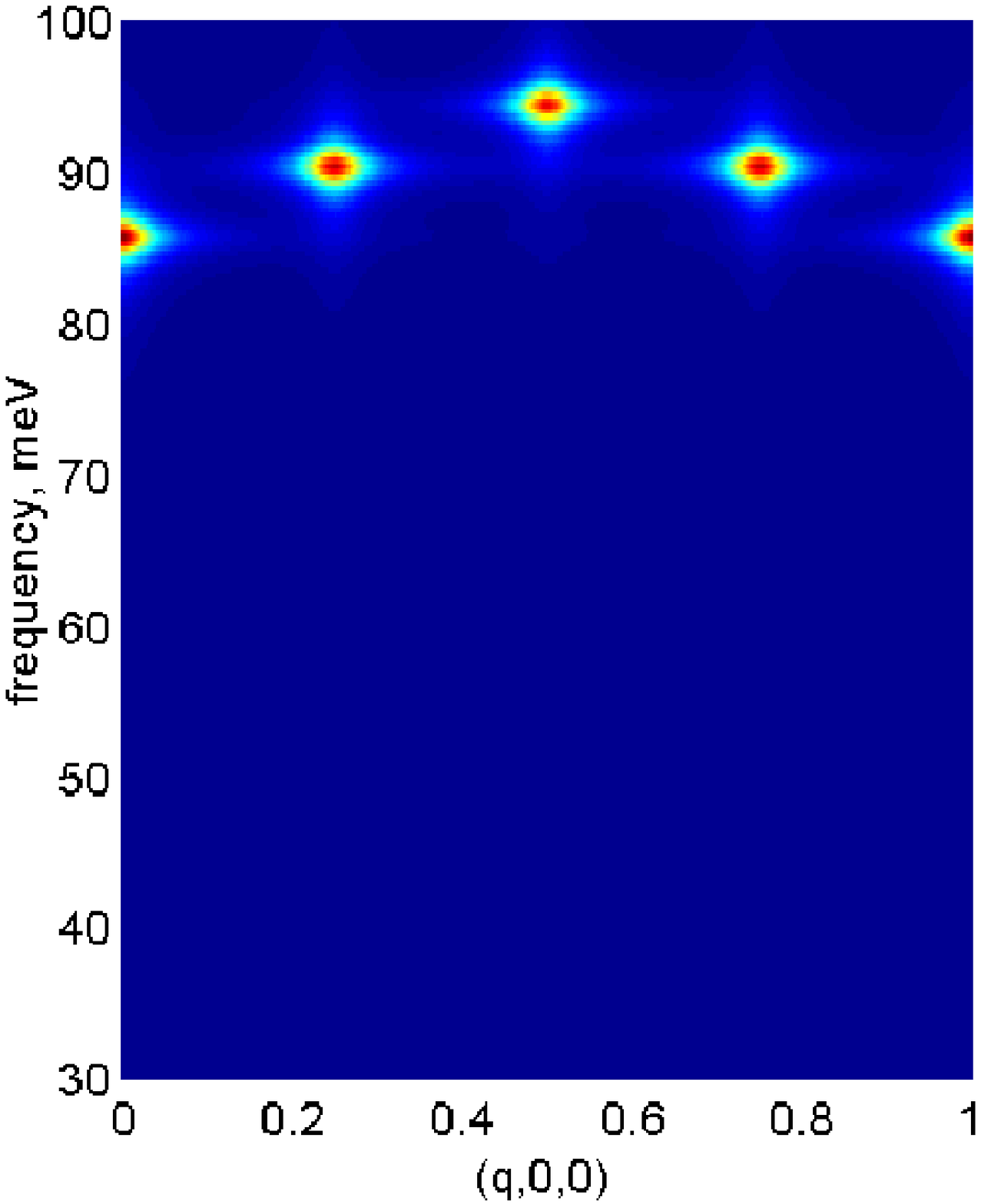} \includegraphics[width
= 1.671 in]{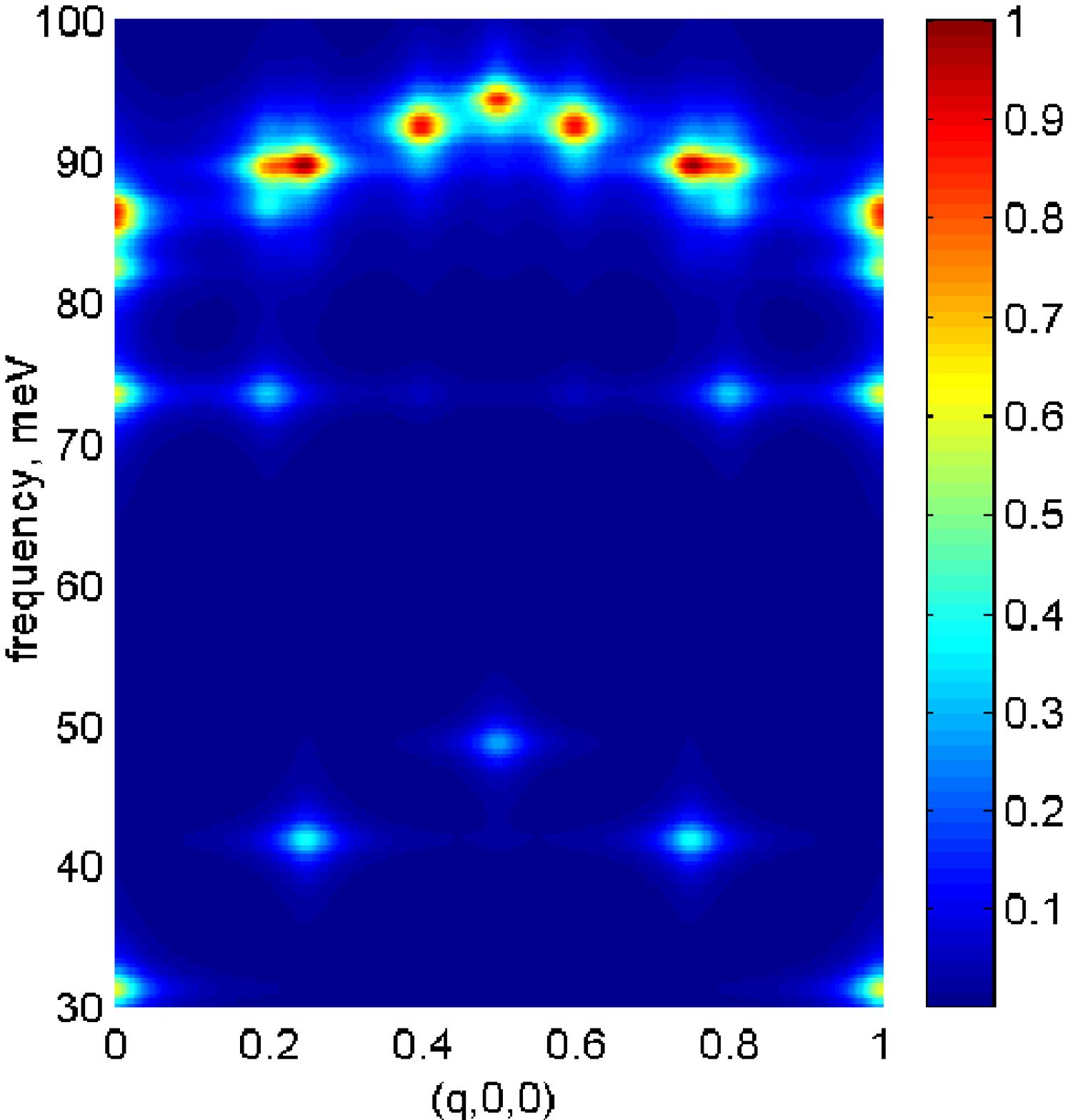}\\
\hspace{.11  cm}\includegraphics[width = 0.96
in]{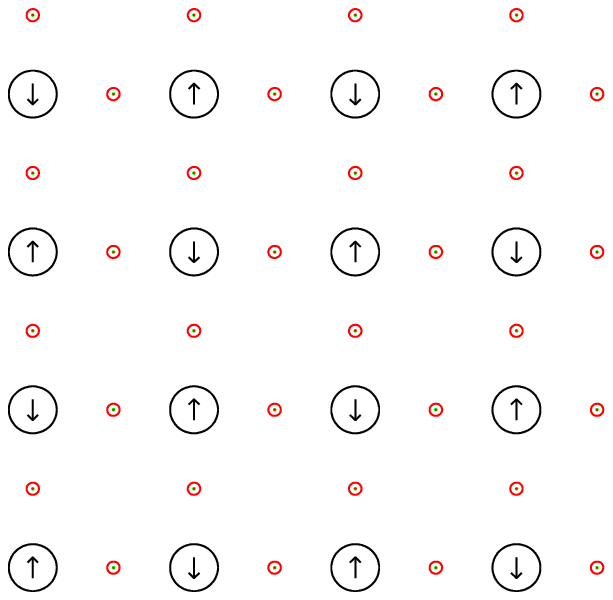}\hspace{.9 cm} \includegraphics[width = 1.2
in]{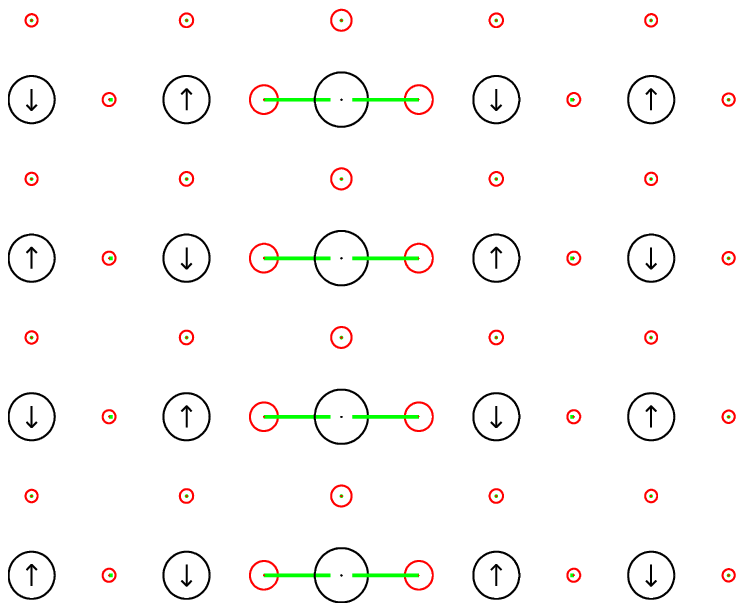}
\vspace{0.2cm} \caption{Neutron scattering spectra and the ground
state structures for the undoped (left panels), and hole-doped ($x =
0.2$) CuO$_2$ planes.  The neutron scattering spectra are symmetrized
with respect to directions along and across the stripe. In the lower
plots, black circles represent Cu sites and red circles represent O.
The radius of the circle is proportional to the corresponding site
hole density.  Arrows centered on circles show the magnitude and
direction of spin.  The green lines originating from O sites indicate
the magnitude and direction of the equilibrium O displacements due to
the formation of a stripe.}
\label{fig:Cu}
\end{center}
\end{figure}

\vspace{-.4cm}
We now turn to the case of doped CuO$_2$ planes for comparison.  To this end,
we analyze the three-band model analogous to Eqs.~(\ref{eq:H0}) and
(\ref{eq:Hc}), but with only one orbital on the metal site (Cu d$_{x^2-y^2}$).
The electron-electron interactions that we include are the Cu on-site repulsion
$U_d = 8$, the O on-site repulsion $U_p = 3$, and the nearest-neighbor Cu-O
repulsion $U_{pd} = 1$.  The remainder of the parameters are
$\epsilon_p-\epsilon_d = 4,\ K = 32/{\AA}$, $\alpha = 4.5/\AA$ and $\beta =
1/\AA$, with $t_{pd} = 1.2$ eV\cite{yuPRB}. Results of calculations for undoped
and doped cuprates are presented in Fig.~\ref{fig:Cu}.

\vspace{-.2cm}

\begin{figure}[htbp]
\begin{center}
\includegraphics[width = 1.35 in]{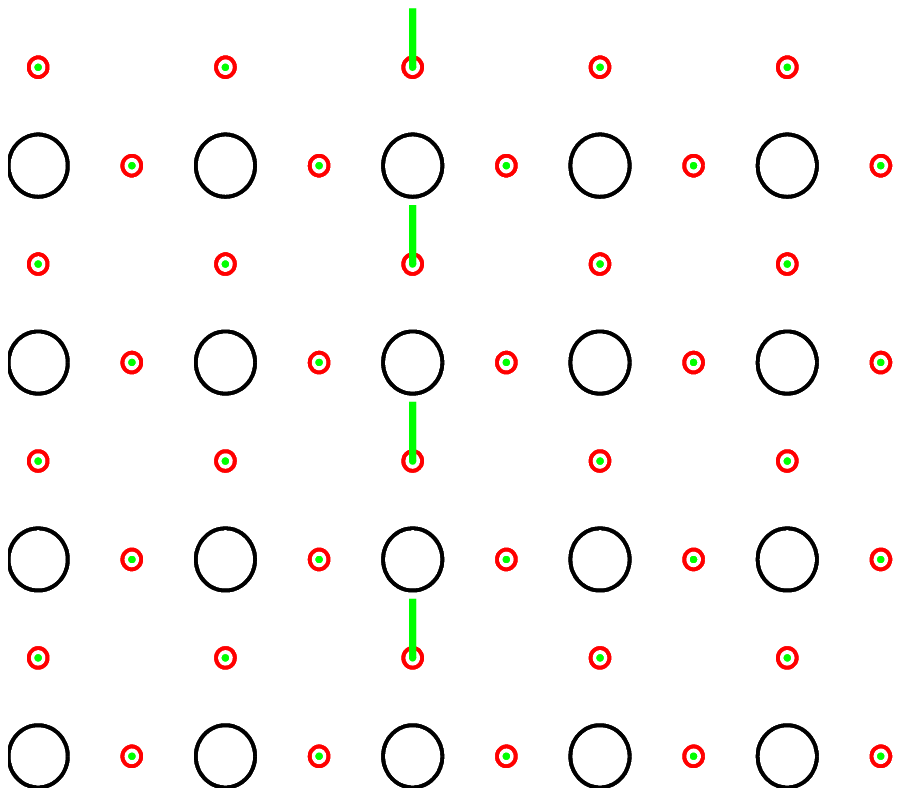}\hspace{.5cm}
\includegraphics[width = 1.4 in]{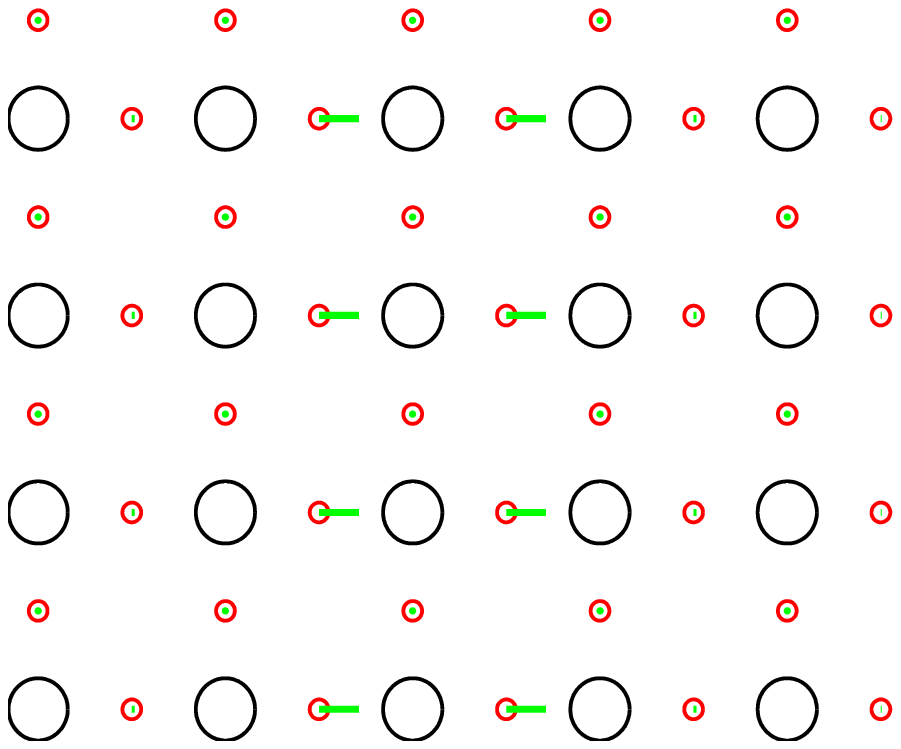}
\vspace{0.4cm} \caption{Localized vibrational eigenmodes for the case of a $x=0.2$ doped CuO$_2$ plane.
The stripe is centered along the middle vertical row of Cu.  Unlike the case of
diagonal stripes in NiO$_2$, there are two localized branches, one that
corresponds to the oxygen vibration parallel to the stripe (low frequency,
$E = 31.2$ meV, on the left), and one that corresponds to the oxygen
displacements perpendicular to the stripe (high frequency, $E = 73.3$ meV, on
the right). }
\label{fig:Cu_mode}
\end{center}
\end{figure}

\vspace{-.5cm}
While generally similar to the NiO$_2$, in the CuO$_2$ case doping produces
$two$ low-frequency branches of localized modes that split off from the band of
extended phonon modes.  Representative modes from each localized branch are
shown in Fig.~\ref{fig:Cu_mode}.  The lower frequency band centered around
$\sim 40$ meV (left) corresponds to the oxygen displacements along the
stripe and the high frequency one at $\sim 73$ meV to the vibrations
perpendicular to the stripe. (The case shown in Fig. 4, right, is a localized
asymmetric breathing mode\cite{robCu1,robCu2}.) Experimental evidence shows
that mode softening of about $15$ percent occurs with doping in the cuprates,
consistent with the softening predicted here for the
$perpendicular$ edge mode\cite{robCu1,robCu2}.  Recent neutron measurements in twinned
and detwinned YBa$_2$Cu$_3$O$_{6+\delta}$
\cite{ybco} have attempted to resolve the phonon spectrum both
parallel and perpendicular to the purported stripe direction\cite{mook}, but
conclusive identification has been hampered by the degeneracy and mixing of
many modes near the zone boundary.  As yet, there has been no search for the
$parallel$ edge modes which are predicted here to have a huge softening of
${\sim 60}$ percent. However, there are distinct anomalies in the longitudinal
phonon spectrum of La$_{2-x}$Sr$_x$CuO$_4$ at 30 meV that have not been studied
in detail\cite{robCu2}.   Also, it is interesting to mention a possible
connection to the ``magnetic resonance'' mode reported at 41 meV \cite{rosat}.
Note that this low-energy mode (see Fig.~\ref{fig:Cu_mode}, left) can only
exist for the vertical stripe and hence is absent for the diagonal stripes in
the case of NiO$_2$ (Fig.~\ref{fig:Ni}).

\vspace{-4 mm}
Despite the impressive agreement between the energy scales of the predicted and
observed modes, it is important to note some deviations from the reported
experimental data in terms of the mode intensities and dispersions.  We
speculate that these discrepancies may be due to the detailed local oxygen
environments: either (a) more complex intra-unit cell distortions, i.e. local
oxygen polarizations (e.g., from local confinement or pairing effects), which
would require more detailed modelling than the constrained O displacements used
here, or (b) more complex stripe configurations than the regular linear stripes
we have considered above. (Such textured stripe patterns are indeed found in
certain models for the origin of stripes\cite{dagotto,branko}).

We conclude that the formation of split-off $local$ modes, well known in the
polaronic physics of conjugated polymers and quasi-1D charge transfer
salts\cite{poly,Raman,RRaman}, is also expected to be prominently manifested in
the case of other doped strongly correlated materials, including nickelates,
cuprates, and bismuthates\cite{1,2,3,4,5,6,7}. Elsewhere we will report the
simultaneous prediction of electron, spin, and lattice edge modes, which should
further facilitate the identification of the intrinsically inhomogeneous phases
by cross correlating electronic gap modes that occur on eV scale with the
localized lattice and spin modes that appear on meV scales.  In addition to
ARPES and inelastic magnetic and lattice neutron scattering, experimental
techniques that could be applied to study the edge modes include infra-red and
Raman spectroscopy\cite{Raman}; resonant Raman scattering would be particularly
valuable even for low doping levels\cite{RRaman}.

We would like to acknowledge helpful discussions with T. Egami.  This work was
supported by the U.S. DOE.

\end{document}